\documentstyle[aps,prl,multicol,epsf]{revtex}
\def\be{\begin{equation}}
\def\ee{\end{equation}}

\def\bea{\begin{eqnarray}}
\def\eea{\end{eqnarray}}
\begin{document}

\title{Formation of Liesegang patterns}
\author{Zolt\'an R\'acz}
\address{Institute for Theoretical Physics,
E\"otv\"os University \\ P\'azm\'any s\'et\'any 1/a, 1117 Budapest, Hungary}

\maketitle

\begin{abstract}
It has been recently shown that precipitation bands 
characteristic of Liesegang patterns emerge from spinodal decomposition 
of reaction products in the wake of moving reaction fronts. 
This mechanism explains the geometric sequence 
of band positions $x_n\sim Q(1+p)^n$ and, furthermore, it 
yields a spacing coefficient $p$ that is in agreement 
with the experimentally observed Matalon-Packter law.
Here I examine the assumptions underlying this theory 
and discuss the choice of input parameters that leads to 
experimentally observable patterns. 
I also show that the so called {\em width law} relating the 
position and the width of the bands $w_n\sim x_n$ 
follows naturally from this theory.
\end{abstract}

\begin{multicols}{2}
\narrowtext

\section{Introduction}

Formation of precipitation patterns in the wake of moving reaction
fronts (known as the Liesegang phenomenon) 
has been studied for more than a century \cite{{liese},{Henisch},{bibli}}. 
The motivation for these studies has been diverse, coming from 
the importance of related practical problems such as crystal 
growth in gels, as well as from the fascination with a 
complex pattern that has eluded a clean-cut explanation 
(e.g. agate rocks are believed to display Liesegang patterns). From a 
theoretical point of view, the main factor in the popularity was
the belief that much can be learned about the details of 
precipitation processes (nucleation, growth, coagulation, etc.)
by investigating the instabilities underlying this phenomenon.
Currently, the Liesegang phenomenon is mainly studied 
as a nontrivial example of pattern formation 
in the wake of a moving front \cite{{dee},{luthi},{us}} and there are  
speculations about the possibility of creating complex 
mesoscopic structures using this rather inexpensive process. 

Liesegang patterns are easy to produce (Fig.1 shows a particular 
experiment that we shall have in mind 
in the following discussion). The main ingredients are 
two chemicals $A$ and $B$  yielding a reaction product $A+B\rightarrow C$
that forms a nonsoluble precipitate $C\rightarrow D$ under appropriate 
conditions [$A=NaOH$, $B=MgCl_2$ and $D=Mg(OH)_2$ in Fig.1]. 
The reagents are separated initially with one of them 
($B$, inner electrolyte) dissolved in
a gel and placed in a test tube. Then at time $t=0$ an aqueous solution
of the other reagent ($A$, outer electrolyte) is poured over the gel. 
The initial concentration $a_0$ of $A$ is chosen to be much larger
than that of $B$ (typically $a_0/b_0\approx 10^2$), thus $A$ diffuses 
into the gel and a reaction front moves down the tube. 
Behind the front, a series of stationary precipitation zones 
(Liesegang bands) appear at positions $x_n$ 
($x_n$ is measured from the interface between the gel and 
the aqueous solution; 
$n=1,2,...,10-20$, typically).
A band appears in a rather short time-interval thus the time of the 
appearance $t_n$ of the $n$-th band is also a well defined, experimentally 
measurable quantity. Finally, the widths of the bands $w_n$ can also 
be determined in order to characterize the pattern in more detail. 
\begin{figure}[htb]
\vspace{1cm}
\centerline{JPEG file attached
           }
\vspace{1cm}
\caption{Liesegang patterns obtained with reagents $A=NaOH$ and $B=MgCl_2$
in polyvinylalcohol gel.
The white precipitate is $D=Mg(OH)_2$. The height of the columns is 30$cm$ 
and it takes about a 1-2 weeks for the patterns to form. The columns show 
different patterns due to the difference in the initial 
concentrations of the outer electrolyte $NaOH$. The experiments were 
carried out by M. Zr\'\i nyi (Technical University of Budapest).}
\end{figure} 

The experimentally measured quantities ($x_n$, $t_n$ and $w_n$) 
in {\em regular} Liesegang patterns satisfy the following 
{\em time-, spacing-,} and {\em width} laws.

\begin{itemize}

\item Time law \cite{morse}: 
\be
x_n \sim \sqrt{t_n}\, .
\label{timelaw}
\ee
\noindent This law is satisfied in all the experiments where it was 
measured and it appears to be a direct consequence of the diffusive 
dynamics of the reagents.
\\
\item Spacing law \cite{jabli}:
\noindent The positions of the bands form a geometric 
series to a good approximation 
\be
x_n \sim Q(1+p)^n
\label{spacing}
\ee
where $p>0$ is 
the spacing coefficient while $Q$ is the amplitude of the spacing law.
The quantitative experimental observations concern mainly this
law. More detailed works go past the confirmation of the 
existence of the geometric series and study the
dependence of the spacing coefficient on $a_0$ and $b_0$. The results 
can be summarized in a relatively simple expression usually referred to as
the {\em Matalon-Packter law}~\cite{{Matalon},{Packter}}:
\be
p=F(b_0) + G(b_0)\frac{b_0}{a_0} 
\label{MatPac}
\ee  
where $F$ and $G$ are decreasing functions of their 
argument $b_0$.
\\
\item Width law \cite{widthlaw}: 
\be
w_n\sim x_n.
\label{wlaw}
\ee
This is the least established law since there are problems with both  
the definition and the measurement (fluctuations) of the width. Recent,
good quality data \cite{width} does support, however, 
the validity of \ref{wlaw}).
\end{itemize}

It should be clear that (\ref{timelaw}-\ref{wlaw}) summarizes 
only those properties of Liesegang patterns that are common in a
large number of experimental observations. There is a wealth of additional 
data on various details such as e.g. the secondary structures
or the irregular band spacing \cite{Henisch}. 
These features, however, appear to be 
peculiarities of given systems. It is hard to characterize them and their
reproducibility is often problematic as well.
In view of this, it is not surprising that the theoretical
explanations of Liesegang phenomena have been mainly concerned with the 
derivation of (\ref{timelaw}-\ref{wlaw}).

The theoretical approaches to quasiperiodic precipitation have a 
long history and the two main lines of thoughts are called as 
pre- and post-nucleation theories (for a brief overview 
see \cite{us}). They all share the assumption that the 
precipitate appears as the system goes through some 
nucleation or coagulation thresholds.
The differences are in the details of treating
the intermediate steps $"...C..."$ in the chain of reactions 
$A+B\rightarrow ... C ... \rightarrow D$ producing the precipitate $D$.
In general, all the theories can explain the emergence of 
distinct bands but only the pre-nucleation theories can account   
\cite{{dee},{luthi},{Wagner},{Prager},{zeldo}}
for the time- and spacing laws of normal patterns.
These theories are rather complicated, however, and have been developed 
only recently \cite{us} to a level that the dependence of $p$ 
on the initial concentrations $a_0$
and $b_0$ can be investigated quantitatively, and connection can be made 
to the Matalon-Packter law \cite{{Matalon},{Packter}}. 

Unfortunately, there are several problems with the theories mentioned above. 
First, they employ a large number of parameters and some of these 
parameters are hard to grasp theoretically and impossible to control 
experimentally (an example is the lower threshold in the density of $C$-s 
below which aggregation $C+D\rightarrow 2D$ ceases \cite{luthi}).
Second, some of the mechanisms invoked in the explanations 
are too detailed and tailored 
to a given system in contrast to the generality of the 
resulting pattern in diverse systems. A real drawback of the
too detailed description is that  
quantitative deductions are difficult to make even with the present 
computer power \cite{dee}. A final problem we should mention is
the absence of an unambiguous derivation of the width law 
in any of the theories.

In order to avoid the above problems, we have recently developed 
a simple model of band formation \cite{ADMRspin} based on the assumption
that the main ingredients of a macroscopic description 
should be the presence of a moving reaction front and the phase 
separation that takes place behind the front. This theory contains 
a minimal number of parameters, it accounts for 
the spacing law, and it is simple enough that the existence of the 
Matalon-Packter law can be established numerically. The apparent success
warrants a closer look at the model and, in this lecture, I will 
describe in detail how one arrives at such a model and what are
the underlying assumptions of the theory. Then I would like to 
discuss the choice of input parameters that yield 
experimentally observable patterns and, finally, I will show that 
the derivation of width law is straightforward in this theory.

\section{The model}

Let us begin building the model by taking a look at 
Fig.1. It shows alternating  high- and low-density regions  
of the chemical $Mg(OH)_2$ and the systems appear to be a quasi-steady
state (actually, there are experiments that suggest that the pattern
does not change over a 30 years period \cite{Henisch}).
We shall take this picture as an evidence that 
phase separation \cite{gun} underlies the formation of bands and, 
furthermore, that the phase separation takes place at a very low 
effective temperature (no coarsening is observed). 

The phase separation, of course, must be preceded by the 
production of $C$-s. This is the least understood part of the 
process and it is particular to each system. What is clear is that 
due to the condition $a_0\gg b_0$ a reaction front 
($A+B\rightarrow something$) moves down the tube diffusively
(note that this is the point where the role of the gel is
important since it prevents convective motion).
The result of the reaction may be rather complex (intermediate products,
sol formation, etc.) and one of our main assumptions is that all these
are irrelevant details on a macroscopic level. Accordingly, the production 
of $C$ will be assumed to be describable by the simplest reaction 
scheme $A+B\rightarrow C$.
  
Once $A+B\rightarrow C$ is assumed, the properties of the front
and the production of $C$-s  
are known \cite{galfi}. Namely, the front moves diffusively with 
its position given by $x_f=\sqrt{2D_ft}$, the production of $C$-s
is restricted to a slowly widening narrow interval [$w_f(t)=w_0t^{1/6}$]
around $x_f$, and the rate of production $S(x,t)$ of $C$-s 
can be approximated by a gaussian (the actual form is not a gaussian, 
see \cite{larralde} for details about a non-moving front)   
\be
S(x,t)=\frac{S_0}{t^{2/3}}\exp\Big[-\frac{[x-x_f(t)]^2}{2w_f^2(t)} 
\Big] \, .
\label{lara}
\ee
The parameter $D_f$ can be expressed through $a_0$, $b_0$, and the
diffusion coefficients of the reagents $(D_a,D_b)$ while $S_0$ and
$w_0$ depends also on the rate constant, $k$, of the reaction 
$A+B\rightarrow C$.

An important property of the front is that it leaves behind a 
constant density $c_0$ of $C$-s \cite{us} and $c_0$ depends 
only on $a_0$, $b_0$, $D_a$ and $D_b$.
This is important because the relevant parameters in the phase
separation are $D_f$ and $c_0$ (where and how much of the $C$-s are 
produced \cite{wcom}) and thus the least available 
parameter $(k)$ does not play a significant role in the pattern formation.  

Having a description of the production of $C$-s, we must now turn to 
the dynamics of their phase separation. Since the emerging pattern
is macroscopic, we shall assume that, on a coarse-grained level, 
the phase separation can be described by the simplest `hydrodynamical' 
equation that respects the conservation of $C$-s. This is the Cahn-Hilliard
equation \cite{CahnHill} or, in other context, it is the equation for 
model B in critical dynamics \cite{HalpHoh}. This equation, however, 
requires the knowledge of the free-energy density (${\cal F}$) of the 
system. For a homogeneous system, ${\cal F}$ must have two minima  
corresponding to the low- ($c_l$) and high-density ($c_h$) states
being in equilibrium (Fig.1). The simplest form of ${\cal F}$ 
having this property and containing a minimal number of parameters is
the Landau-Ginzburg free energy (Fig.2)
\be
{\cal F}=-\frac{1}{2}\varepsilon m^2 + \frac{1}{4}\gamma m^4 +
\frac{1}{2}\sigma (\nabla m)^2 \, ,
\label{LGfree}
\ee
where $m=c-(c_l+c_h)/2$ is the density, $c$, of the $C$-s 
measured from the average 
of the two steady state values (we are following the notation in 
\cite{ADMRspin} where the `magnetic language' has its origin in a 
connection to Ising lattice gases). The parameters $\varepsilon$, 
$\gamma$, and $\sigma$ are system dependent 
with $\varepsilon > 0$ ensuring that the system is 
in the phase-separating regime, $\sigma >0$ provides stability
against short-wavelength fluctuations, and requiring 
$\sqrt{\varepsilon/\gamma}=(c_h-c_l)/2$ fixes the minima of ${\cal F}$
at $\pm m_e$ corresponding to $c_l$ and $c_h$. Note that the 
$m\rightarrow -m$ symmetry is usually not present in a real system and 
${\cal F}$ could contain e.g. an $m^3$ term. The presence or absence of 
the $m\rightarrow -m$ symmetry, however, is not relevant for the 
discussion that follows.   
 
\begin{figure}
\centerline{
        \epsfysize=6cm
        \epsfbox{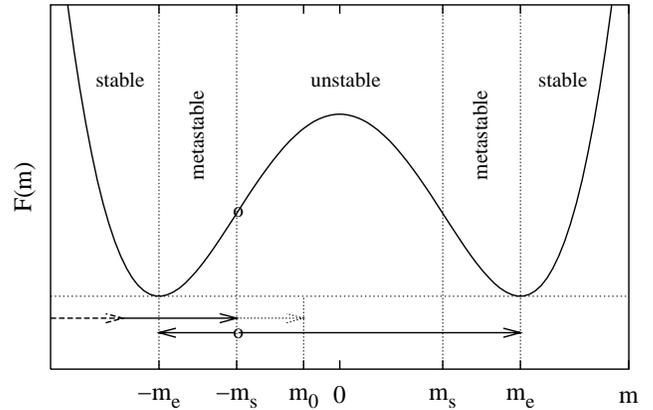}
           }
\vspace{0.5cm}

\caption{The homogeneous part of the free energy as a function of 
$m=c-(c_l+c_h)/2$. The phase separation is an activated process in the
metastable regimes while it goes by spinodal decomposition in the 
(linearly) unstable regime. The spinodal point ($\circ$) 
separates these regimes. The arrows are meant to illustrate how the density
in the front increases towards $m_0$ and how the phase separation to 
the steady states ($\pm m_e$) takes 
place when the density reaches the spinodal value $-m_s$.}
\end{figure} 

Using (\ref{LGfree}) and including the source term, the Cahn-Hilliard 
equation takes the form
\be
\partial_t m= -\lambda \Delta\left(\,\epsilon m-\gamma m^3 +\sigma 
\Delta m\,\right) + S \, .\label{move}
\ee
were $\lambda$ is a kinetic coefficient. The above equation should 
contain two noise terms. One of them should be the
thermal noise while the other should originate in the chemical reaction
that creates the source term.  
Both of these noise terms are omitted here. 
The reason for neglecting the thermal noise is the low
effective temperature of the phase separation as discussed in 
connection with Fig.1. The noise in $S$, on the other hand, is dropped 
since the $A+B\to C$ type reaction fronts have been shown to be 
mean-field like above dimension two \cite{cornell}.

The absence of noise means that the phase separation can occur only 
through spinodal decomposition \cite{{gun},{spindec}}. Thus the assumption 
behind omitting the noises is that the characteristic time of 
nucleation is much larger than the time needed by the front to increase 
the density of $C$-s beyond the spinodal value ($-m_s$ in Fig.2) where 
the system is unstable against linear perturbations. Since there are 
examples where the bands appear to be formed by nucleation and growth
\cite{Henisch}, the 
spinodal decomposition scenario is clearly not universally applicable,
and one should explore the effects of including noise (this becomes, 
however, an order of magnitude harder problem).

Eq.(\ref{move}) together with the form of the source (\ref{lara}) 
defines now our model \cite{ADMRspin} that produces regular 
Liesegang patterns (Fig.3) satisfying the spacing law (\ref{spacing}) and, 
furthermore, the spacing coefficient is in agreement 
with the Matalon-Packter law (\ref{MatPac}). Fig.3 shows a rather general
picture that is instructive in understanding the pattern formation. The
last band acts as a sink for neighboring particles above $-m_e$ ($c_l$) 
density. Thus the $C$-s produced in the front end up increasing the 
width of the last band. This continues until the front moves far enough
so that the density in it reaches the spinodal value. Then the 
spinodal instability sets in and a new band appears. Remarkably, 
the above picture is rather similar to the phenomenological `nucleation 
and growth' scenario \cite{us} with the density at the spinodal 
point playing the role of threshold
density for nucleation. It is thus not entirely surprising that both of
these theories do equally well in producing the spacing- and the 
Matalon-Packter law.

One should note that the actual form of ${\cal F}$ does not play an 
important role in the picture developed above. The crucial feature is 
the existence of a spinodal density above which phase separation occurs.
This is the meaning of our previous remark about the irrelevance 
of the $m^3$ term in the free energy (of course, one should also 
realize that explanations of 
details in experiments may require the inclusion of such terms).

\begin{figure}
\centerline{
        \epsfysize=6cm
        \epsfbox{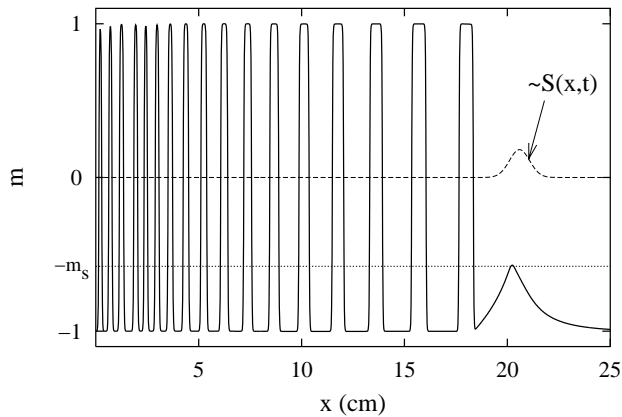}
           }
\vspace{0.5cm}
\caption{Liesegang pattern obtained for
front parameters $D_f=21.72$, $w_0=4.54$, and $S_0=0.181$ with
length, time, and $m$ (concentration)
measured in units of $\sqrt{\sigma/\epsilon}=2\cdot 10^{-4}$m,
$\sigma/(\lambda \epsilon^2)=40$s, and $\sqrt{\epsilon/\gamma}$, respectively.
The dashed line denotes the rate of production of $C$-s
($S$), measured in units of
$\lambda\epsilon^{5/2}/(\gamma^{1/2}\sigma)$ and magnified by a factor
$2\cdot 10^5$. The dotted line represents the density at
spinodal point, $-m_s=-1/\sqrt{3}$.}
\end{figure} 


\section{Choice of parameters}

Fig.3 shows the results of numerical solution of eq.(\ref{move}) with 
the same parameter values as in Fig.3 of ref.\cite{ADMRspin} but stopped 
at an earlier time so that the visual similarity to the experiments 
(number of bands in Fig.1) would be greater. In this section, we shall 
examine whether the parameters used for obtaining this resemblance 
have any relevance to real Liesegang phenomena. 

The experimental patterns have a total length of about 
$\ell_{exp}\approx 0.2$m and the time of producing such a pattern is about
1-2 weeks (we shall take $\tau_{exp}\approx 10^6$s). Since our model 
has a length-scale $\ell_{th}=\sqrt{\sigma/\varepsilon}$ and a time-scale 
$\tau_{th}=\sigma/(\lambda\varepsilon^2)$, they can be chosen so 
[$\sqrt{\sigma/\epsilon}=2\cdot 10^{-4}$m and 
$\sigma/(\lambda \epsilon^2)=40$s] that $\ell_{exp}\approx \ell_{th}$ and
$\tau_{exp}\approx \tau_{th}$. Once we have chosen $\ell_{th}$ and 
$\tau_{th}$ we can start to calculate other quantities and see if they
have reasonable values.

It is clear from Fig.3 that the widths of the bands are in agreement 
with the experiments, they are of the order of a few mm at the beginning
and approach to $\sim1$cm at the end. The width of the front is also 
of the order of 1cm after $10^6$s. Unfortunately, 
there is no information on the reaction
zone in this system. In a study of a different system
\cite{Kopelman} it was found that $w_f$(t=2 hours)$\approx$ 2mm. Extrapolating 
this result to $t=10^6$s one finds $w_f\approx 1$cm [note that the exponent
of the increase of $w_f(t)\sim t^{1/6}$ is small] in agreement 
with the observed value.

Next we calculate the diffusion coefficient of the front, 
$D_f=21.72\cdot \ell_{th}^2/\tau_{th}\approx 2\cdot 10^{-8}$m$^2$/s. 
This value appears to be an order of magnitude larger than the usual 
ionic diffusion coefficients ($D\approx 10^{-9}$m$^2$/s). One should 
remember, however, that Fig.3 is the result for initial conditions 
$a_0/b_0=10^2$ \cite{ADMRspin} and, for this ratio of $a_0/b_0$, the 
diffusion coefficient of the front $D_f$ is about 10 times larger 
than $D_a$ ($D_f/D_a\approx 10$ see Fig.4 in \cite{us}). Thus $D_f$ 
also comes out to be the right order of magnitude.

We do not have information on the amplitude ($S_0$) of the source but,
once the concentrations ($a_0$, $b_0$) are given and $D_f$ and $w_f$
are known then $S_0$ is fixed by the conservation law for the $C$-s.
Thus the correct order of magnitude for $D_f$ and $w_f$ should
ensure that $S_0$ is also of right order of magnitude.

\begin{figure}
\centerline{
        \epsfysize=6cm
        \epsfbox{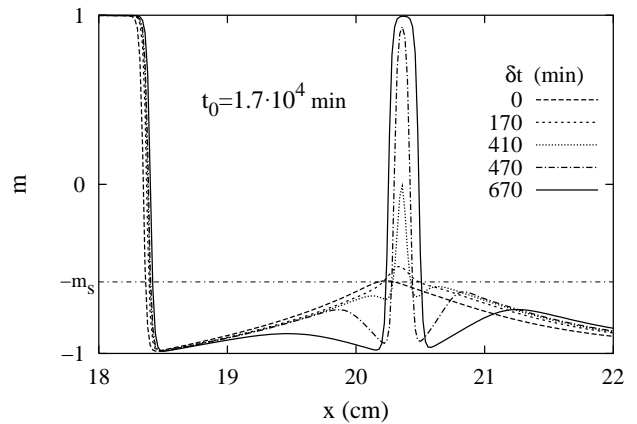}
           }
\vspace{0.5cm}
\caption{Details of the time-evolution of the formation of the
last band in Fig.3. The time $t_0$ is the moment 
when the concentration at the front reached 
the spinodal value and $\delta t$ is measured from $t_0$.}
\end{figure} 

Finally, we shall calculate the time it takes
for a band to form. It is well known that the bands appear rather 
quickly. From the visual notice of the beginning of the 
band formation it takes about $\tau_{ini}=30-60$ minutes for the band 
to be clearly seen and then it takes much longer to increase its width
to the final value.
In order to calculate $\tau_{ini}$ let us consider the formation 
of the last band in Fig.3 (see Fig.4). The lower limit of the 
density that can be visually noticed is, of course, not well defined. 
We shall assume that this density
corresponds to $m=0$ i.e. it is the halfway density from $c_l$ to $c_h$.
This means that we see the beginnings of the band at $\delta t=410$min and 
the density reaches well above $90$\% of its final value by 
$\delta t=470$min. Consequently, we obtain again an estimate 
($\tau_{ini}\approx 60$min) for an observed quantity that agrees with
the experiments.
As a result of the above estimates, we feel that the parameters 
in our model can indeed be chosen 
so that they are relevant to real Liesegang experiments.


\section{Width law}

The width law is problematic  from experimental point of view since
the fluctuations in the widths appear to be large. Part
of the difficulties are undoubtedly due 
to the fact that the boundaries of the bands are not 
sharply defined and high-resolution digitizing methods are needed 
in a precise analysis. 
The most thorough experiment to date has been carried out recently 
\cite{width} with the result $w_n\sim x_n^\alpha$ 
where $\alpha \approx 0.9-1.0$.

As to the theories, they also have their share of difficulties since,
on a microscopic level, the growth of the width involves precipitation 
processes in the presence of large concentration gradients, while a
macroscopic treatment must elaborate on the dynamics of the
interfaces between two phases. Accordingly, there are only a few works 
to report on. Dee \cite{dee} used reaction-diffusion equations 
supplemented by terms coming from nucleation and growth processes
and obtained $w_n\sim x_n$ from a rather limited (6 bands) numerical 
result. Chopard et al. \cite{luthi} employed cellular automata simulations
of a phenomenological version of the microscopic processes and
found $w_n\sim x_n^\alpha$ with $\alpha \approx 0.5-0.6$. Finally,
Droz et al. \cite{width} combined scaling considerations with 
the conservation law for the number of $C$ particles to obtain 
$\alpha$ in terms of the scaling properties of the density of precipitates
in the bands. Assuming constant density they found $\alpha=1$.
Our derivation below parallels this last work in that the same 
conservation law is one of the main ingredient in it.

In our theory, the derivation of the width law is straightforward.
One combines the facts that
(i) the reaction front leaves behind a constant density ($c_0$) of $C$-s,
(ii) the $C$-s segregate into low ($c_l$) and high ($c_h$) density bands,
(iii) the number of $C$-s is conserved in the segregation process;  
and writes down the equation expressing the conservation of $C$-s
\be
(x_{n+1}-x_n)c_0=(x_{n+1}-x_n-w_n)c_l +w_nc_h \, .
\label{widtheq}
\ee
Using now the spacing law (\ref{spacing}) that has been established 
for this model one finds
\be 
w_n=\frac{p(c_0-c_l)}{c_h-c_l}x_n=\zeta x_n \, .
\label{widthfinal}
\ee

We have thus derived the width law and obtained the coefficient 
of proportionality, $\zeta$, as well. The importance of $\zeta$ 
lies in that measuring it provides
a way of accessing $c_0$ that is not easily measured otherwise.

\section{Final remarks}

In summary, we have seen that the spinodal decomposition scenario 
for the formation of Liesegang patterns performs well whenever 
quantitative comparison with experiments is possible.
It remains to be seen if the applicability of this model 
extends beyond the {\em regular} patterns. One should certainly try 
to use this theory to explain the {\em exotic} patterns 
(e.g. inverse patterns, helixes) that are 
experimentally reproducible and lack even qualitative understanding.

\section*{Acknowledgments}
I thank M. Droz, M. Zr\'\i nyi, T. Antal, P. Hantz, J. Magnin, and T. Unger 
for useful discussions.
This work has been supported by the Hungarian Academy of Sciences (Grant No.
OTKA T 029792).

\end{multicols}


\begin{references}
\bibitem{wcom} For usual parameter values, the width is much smaller 
               than all the other length-scales and its actual value 
               does not significantly affect the onset of phase 
               separation.
\bibitem{spindec} The spinodal decomposition takes place in the 
                  regime where the homogeneous solutions of eq.(\ref{move})
                  are linearly unstable.
\end{references}
\end{document}